# Experimental Realization of a Quantum Spin Pump

Susan K. Watson*†, R. M. Potok†, C. M. Marcus†, and V. Umansky‡

*Department of Physics, Middlebury College, Middlebury, VT 05753, USA

†Department of Physics, Harvard University, Cambridge, MA 02138, USA

‡Braun Center for Submicron Research, Weizmann Institute of Science, Rehovot 76100, Israel

**We demonstrate the operation of a quantum spin pump based on cyclic radio-frequency excitation of a GaAs quantum dot, including the ability to pump pure spin without pumping charge. The device takes advantage of bidirectional mesoscopic fluctuations of pumped current, made spin-dependent by the application of an in-plane Zeeman field. Spin currents are measured by placing the pump in a focusing geometry with a spin-selective collector.**

Using electron spin to encode information in semiconductors holds promise for integrating computation and storage [1] and, in coherent systems, is expected to provide significantly increased immunity from environmental decoherence compared with conventional charge-based electronics [1, 2]. Among the needed elements for any spin-based electronic system is a device that generates a spin current, the analog of a battery in conventional electronics. Candidates for such devices include injection schemes based on magnetic semiconductors [3, 4] and ferromagnetic metals [5, 6], ferromagnetic resonance devices [7], and a variety of spin-dependent pumps [8-14].

In this Letter, we demonstrate the operation of such a quantum-dot–based spin pump—including the ability to pump pure spin without pumping charge—using a gate-defined lateral quantum dot fabricated on a GaAs/AlGaAs heterostructure. Pumping of charge using cyclic gate voltages applied to a phase-coherent dot has been the subject of numerous investigations (mostly theoretical) in the past several years [15-20]. Quantum pumps are closely related to classical charge pumps using Coulomb blockade [21, 22] as well as higher-frequency mesoscopic photovoltaic effects [20, 23-25] and photon-assisted tunneling [26]. A recent proposal [8] (MCM) considered a quantum-dot-based charge pump in the presence of sizable Zeeman splitting, and showed that such a device would function as a phase-coherent spin pump. An important feature of the MCM proposal is that it remains operational regardless of whether the pumped current arises from adiabatic pumping [15-18, 20], mesoscopic rectification [18, 27, 28], or photovoltaic mechanisms [20, 23] as long as the fluctuations of pumped current are larger than the average, so that both positive and negative current can be generated and controlled by external parameters such as device shape or applied magnetic flux.

In order to realize the spin pump device experimentally, radio-frequency (rf) sinusoidal voltages were applied to two shape-defining gates of a quantum dot, producing a dc current through the quantum dot due to a combination of adiabatic pumping and rectification effects. As discussed in detail below, while pumped charge can be directly measured directly across the device, measuring the pumped *spin* is more subtle, and in the present set-up is detected using an electron focusing configuration with a quantum point contact operating as a spin detector [29, 30].

Due to quantum coherence, the direction of the pumped current is a mesoscopically fluctuating quantity with zero average. In the absence of an external magnetic field, $B_\parallel = 0$, the two spin states are degenerate and the pumped currents for spin up ($I_\uparrow$) and spin down ($I_\downarrow$) are identical, fluctuating together as control parameters are swept (this is also the case for mesoscopic rectification [28]). In this zero-field case, pumping induces a net charge current $I_c = I_\uparrow + I_\downarrow \neq 0$, but no net spin current, $I_s = I_\uparrow - I_\downarrow = 0$, as illustrated schematically in Fig. 1a. For sufficiently strong in-plane magnetic fields, $g\mu_B B_\parallel > (kT, \Gamma)$, spin-up and spin-down pumped currents are uncorrelated. Here, $\Gamma$ is the level broadening due to escape and dephasing, $g \sim -0.4$ is the electron g factor, and $\mu_B$ is the Bohr magneton. In this high-field case, pumped charge is comprised of independent contributions of $I_\uparrow$ and $I_\downarrow$ (Fig. 1b), and in general a nonzero spin current exists. In practice, it is straightforward in this regime to tune the charge current to zero (using gate-defined dot shape or a small applied perpendicular magnetic flux) resulting in a *pure spin current*. The variance of the pumped charge (spin) current, $I_{c,s}$, is given by $\overline{I_{c,s}^2} = \overline{I_\uparrow^2} + \overline{I_\downarrow^2} \pm 2\overline{I_\uparrow I_\downarrow} = 2\left[\overline{I_\uparrow^2} \pm \overline{I_\uparrow I_\downarrow}\right]$. At large $B_\parallel$, where $I_\uparrow$ and $I_\downarrow$ become uncorrelated $\left[\overline{I_\uparrow I_\downarrow} \approx 0\right]$, the typical spin current $I_s$ becomes nonzero and the variance of $I_c$ decreases by a factor of two compared to its low field value, where $\overline{I_\uparrow I_\downarrow} \approx \overline{I_\uparrow^2}$ (neglecting the effects of time-reversal symmetry and spin-orbit coupling [13]).

To detect the presence of a spin current, we take advantage of a spin-sensitive electron focusing geometry (inset, Fig. 2), which allows the pumped current to be focused into a collector quantum point contact (QPC), as shown in Fig. 2. In moderate in-plane fields, $B_\parallel \sim 3T$, the collector QPC has been shown to act as a spin-sensitive detector whenever its conductance $g_c$ is tuned (by gate voltages) to $g_c \approx 1e^2/h$ [29, 30]. In this spin-sensitive regime, the base-collector voltage reflects the polarization of the current impinging upon the collector QPC. As a control, when the collector QPC is non-spin-sensitive, which is achieved—even in high fields—by setting $g_c \approx 2e^2/h$, the base-collector voltage signal reflects the total charge current impinging on the collector QPC.

The complete system, comprising the quantum-dot spin pump plus the QPC-focusing test structure, was fabricated on a GaAs/AlGaAs 2DEG using e-beam patterned CrAu depletion gates and nonmagnetic (PtAuGe) ohmic contacts. The high-mobility material $(\mu \approx 5.5 \times 10^6 \, cm^2 V^{-1} s^{-1})$ was useful for obtaining good focusing but is not necessary for the operation of the spin pump itself. This mobility, and sheet density $n \approx 1.3 \times 10^{11} \, cm^{-2}$, gave a mean free path of $45 \, \mu m$. The quantum dot, which has an area of $A \sim 0.1 \, \mu m^2$, is typically operated with one fully open channel in each of its point contact leads, $g_{lead1} \sim g_{lead2} \sim 2e^2/h$.

Focusing was first tested and calibrated using a current-bias configuration $(I_{bias} \approx 1 nA)$. Results are shown in Fig. 2. Next, two sinusoidal signals at 10 MHz with controllable phase difference, $\phi$, were applied (via synchronized Agilent 33250 synthesizers) to two of the confining gates of the dot: $V_{gate1}(t) = V_{dc1} + V_1 \sin(\omega t)$, $V_{gate2}(t) = V_{dc2} + V_2 \sin(\omega t + \phi)$. Applying ac gate voltages of 70 mV (comparable to the characteristic gate-voltage scale of gate-induced mesoscopic conductance or pumping fluctuations (see Fig. 3)) induced a dc current through the dot on the scale of ~ 10–100 pA, measured using an Ithaco 1211 current amplifier with input impedance ~20 $\Omega$ (much lower than the ohmic contact resistance of ~1k$\Omega$.) The rf applied to the gates was modulated at 11 Hz to allow lock-in detection of both the pumped current, measured through the base-emitter circuit, and the voltage across the collector QPC.

The sinusoidal dependence of pumped current on $\phi$, $I(\phi) \sim I_o \sin(\phi)$ (Fig. 3a, inset) is consistent with both adiabatic quantum pumping [17, 19] and capacitively coupled rectification [28]. However, the fact that the magnitude of current is typically larger than one electron per cycle (=1.6 pA at 10 MHz) suggests that the pumped current is dominated by rectification rather than adiabatic pumping. Again, this does not affect the performance of the spin pump.

Figure 3 shows pumped current and collector voltage at $B_\parallel = 4T$ for both spin-selective and nonselective settings of the collector QPC, as a function of dc voltage on one of the shape-defining gates of the quantum dot [31]. The point contacts of the dot were each set at $2e^2/h$, the pumping amplitude was $V_1 = V_2 = 70 mV$ and the phase was $\phi = 3\pi/2$. These data illustrate our main experimental observation: when the collector QPC is set to $g_c = 2e^2/h$, and hence is *not* spin selective, the collector voltage closely follows the pumped current. In contrast, when the collector QPC is spin selective, $g_c = 1e^2/h$, the collector voltage appears unrelated to the pumped total charge current. The fact that only the spin-selective detector shows a signal unrelated to the total current means that a spin current, *fluctuating independently from the total current*, is being generated by the pump. One may readily identify in Fig. 3 several zeros of the total current and see that the spin currents measured at these values of $V_{dc1}$ are nonzero. At these points, the pump operated as a pure spin pump, producing tens of spins per cycle, with zero net charge being pumped.

The corresponding control experiment at $B_\parallel = 0$ (Fig. 4) again compares collector voltages measured at $g_c = 2e^2/h$ and $g_c = 1e^2/h$. In this case, there is little dependence of the collector voltage on the collector QPC conductance. This behavior is expected when the collector QPC is not spin selective and the quantum dot pump produces no spin polarized current.

In summary, we have demonstrated a mesoscopic spin pump using an ac driven phase-coherent quantum dot in a Zeeman field. Spin current—including pure spin current, without any charge current—is detected using a spin-sensitive focusing technique [29, 30]. While this experiment required the application of a sizeable in-plane magnetic field, one can expect similar results using permanent magnets microfabricated along with the quantum dot. This would be particularly effective in materials with a larger g factor than GaAs. It is of interest to clarify how spin-orbit coupling affects the operation of the pump [13]. We speculate that in strong spin-orbit materials it may be sufficient to break time-reversal symmetry with a small applied field (on the scale of a few flux quanta though the dot, typically of order 0.01T for a square-micron dot area). These interesting extensions await further experimental investigation.

We thank Joshua Folk for valuable discussions and assistance in carrying out the measurements, and Piet Brouwer for useful discussions. Research at Harvard was supported by the DARPA-SpinS program and the National Science Foundation under DMR 0072777. R. Potok acknowledges support of an ARO-QC Fellowship though the Army Research Office. S. Watson acknowledges support from Middlebury College and the National Science Foundation (DMR-0074930).

[31] To compensate for a small, unintentional cross-coupling between the collector QPC voltage and the gates of the emitter-dot, the current data shown in Figs. 3 and 4, measured at various collector gate voltages, are shifted horizontally (by < 20mV) to bring the traces into alignment. Collector voltage traces, reflecting either spin polarization or total current, depending on setting of the collector QPC, are then shifted horizontally by the same amount as their corresponding current trace. Similarly, the collector voltage exhibits a smooth background that shifts linearly with $V_{dc1}$ of the emitter dot due to capacitive cross-coupling. This smooth background has been subtracted from traces in Figs. 3 and 4.

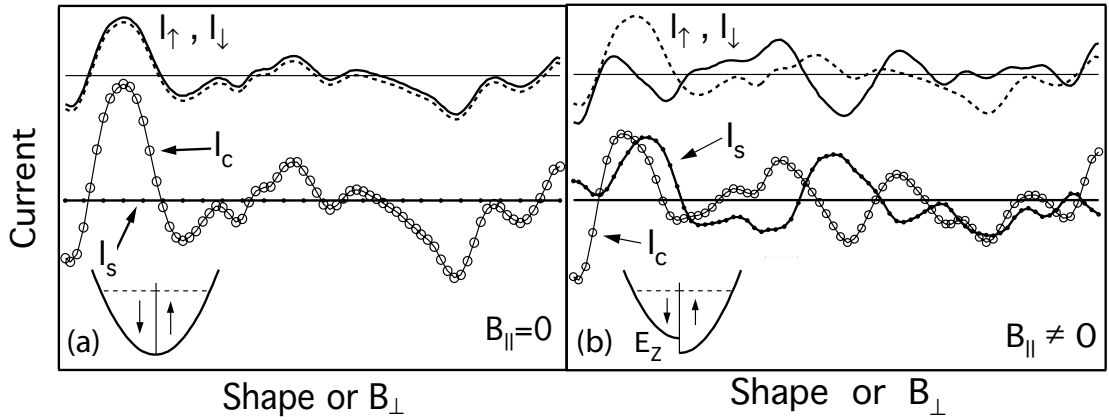

**Figure 1.** Schematic illustration of how mesoscopic pumping fluctuations plus Zeeman splitting can give rise to independent fluctuations of charge current $I_c = I_\uparrow + I_\downarrow$ and spin current $I_s = I_\uparrow - I_\downarrow$, based on Ref. 8. (a) At $B_\parallel = 0$ one has $I_\uparrow = I_\downarrow$ and hence $I_s = 0$. (b) At large in-plane fields, $B_\parallel > (kT, \Gamma)$, (see text) spin degeneracy is lifted, $I_\uparrow$ is uncorrelated with $I_\downarrow$, and in general $I_s \neq 0$. In this case, one may even have a pure spin pump when $I_s \neq 0$ while $I_c = 0$.

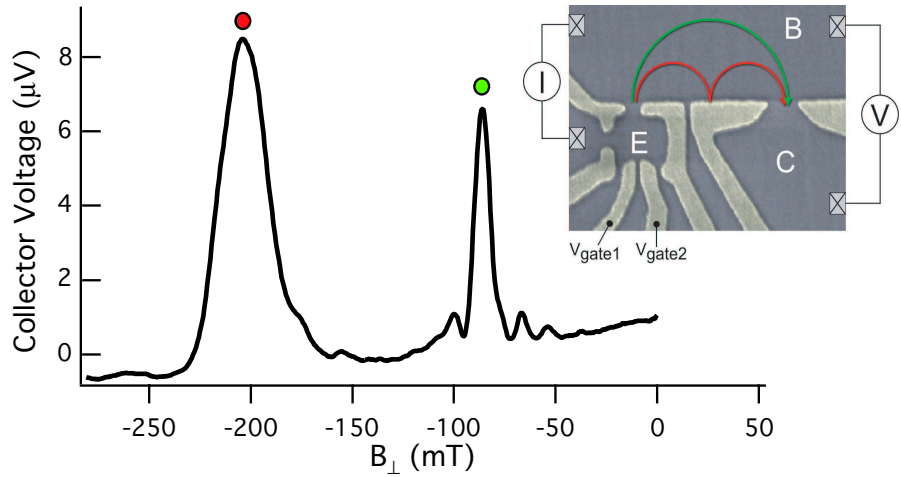

**Figure 2.** The base-collector voltage as a function of perpendicular magnetic field, $B_\perp$. The first (green dot) and second (red dot) focusing peaks in the base-collector voltage occur when the spacing between the emitter (quantum dot) and the collector (QPC) is a multiple of the cyclotron diameter. To find the focusing peaks, a current bias of 1 nA was applied across the dot. Inset: Electron micrograph of the dot/focusing device, with added circuit and schematic trajectories. Electrons emitted from the quantum dot emitter, "E,,, follow ballistic trajectories through the base region, "B,,, into the collector, "C,,. Ac voltages $V_{gate1}$ and $V_{gate2}$ are applied to shape-defining gates.

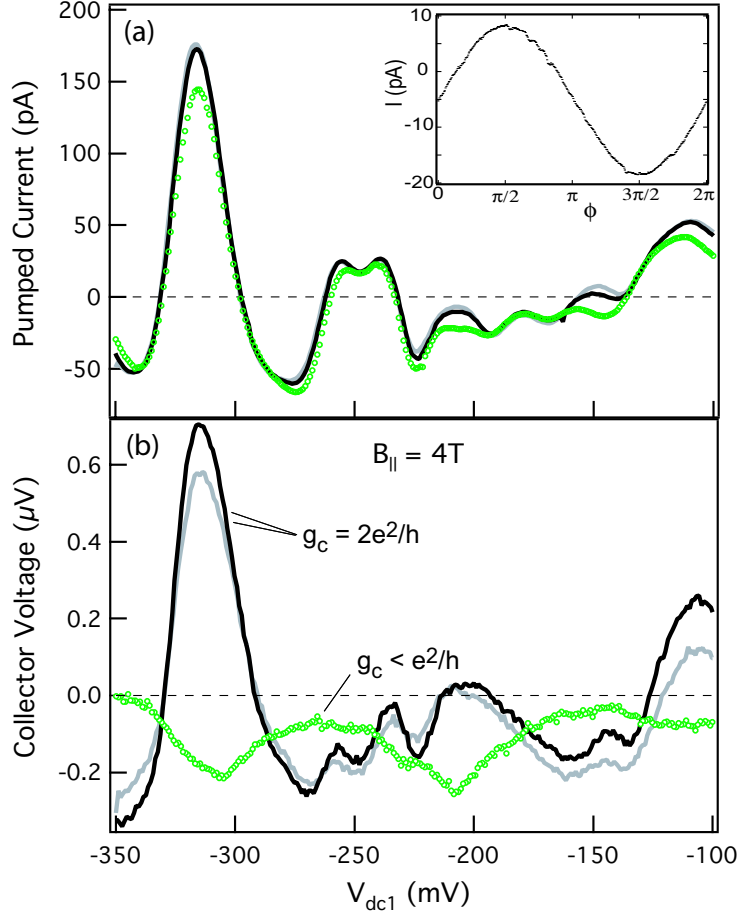

**Figure 3.** (a) The pumped charge current at $B_\parallel \approx 4T$ and $\phi \approx 3\pi/2$ as a function of dot shape, controlled by dc gate voltage $V_{dc1}$. Traces at slightly different collector QPC settings, both with $g_c \approx 2e^2/h$ (gray and black) are similar, and are also similar to the trace with $g_c \approx 1e^2/h$ (green). This is not surprising, as the collector QPC should not affect pumping by the emitter. Inset: Pumped current as a function of the phase $\phi$ between the $V_{gate1}$ and $V_{gate2}$. (b) Collector voltages (relative to the base region) on the second focusing peak under the same conditions as corresponding traces in (a). For spin-nonselective collector QPC, $g_c \sim 2e^2/h$, (gray and black) the collector voltages track pumped current, indicating that the voltage measures the total charge current. For spin-selective collector QPC $g_c \approx 1e^2/h$ (green), the collector voltages are distinctly different from the pumped current and from traces at $g_c \sim 2e^2/h$, indicating a nonzero spin current.

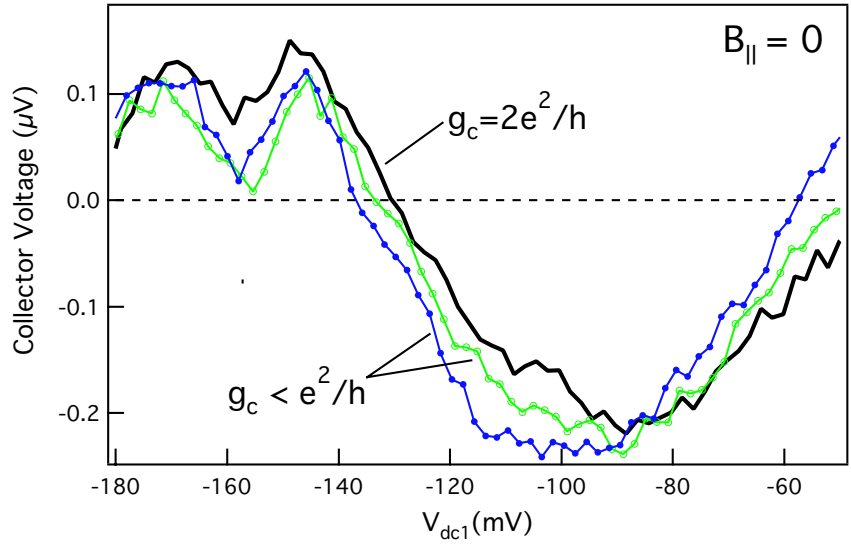

**Figure 4.** Collector voltage as a function of dot shape for $B_\parallel \approx 0$ at two different collector QPC settings with $g_c \approx 1e^2/h$ and one with $g_c \sim 2e^2/h$. Unlike the situation at $B_\parallel \approx 4T$, the collector voltage is not sensitive to the conductance of the collector QPC. This is anticipated as there is no spin selectivity of the collector QPC nor is there a pumped spin current from the dot.